\documentclass[preprint]{aastex631}

\received{$\bigstar$}
\revised{$\bigstar$}
\accepted{$\bigstar$}

\shorttitle{Non-Radial Flows Inside CMEs: STEREO/PLASTIC}
\shortauthors{Al-Haddad et al.}

\begin{document}

\title{Investigating The Cross-section of Coronal Mass Ejections Through the Study of Non-Radial Flows with STEREO/PLASTIC}

\correspondingauthor{Nada Al-Haddad}
\email{nada.alhaddad@unh.edu}

\author[0000-0002-0973-2027]{Nada Al-Haddad}
\affiliation{Space Science Center, Institute for the Study of Earth, Oceans, and Space, and Department of Physics and Astronomy, University of New Hampshire, USA}

\author[0000-0003-3752-5700]{Antoinette B. Galvin}
\affiliation{Space Science Center, Institute for the Study of Earth, Oceans, and Space, and Department of Physics and Astronomy, University of New Hampshire, USA}
\author[0000-0002-1890-6156]{No{\'e} Lugaz}
\affiliation{Space Science Center, Institute for the Study of Earth, Oceans, and Space, and Department of Physics and Astronomy, University of New Hampshire, USA}
\author[0000-0001-8780-0673]{Charles J. Farrugia}
\affiliation{Space Science Center, Institute for the Study of Earth, Oceans, and Space, and Department of Physics and Astronomy, University of New Hampshire, USA}
\author[0000-0002-2917-5993]{Wenyuan Yu}
\affiliation{Space Science Center, Institute for the Study of Earth, Oceans, and Space, and Department of Physics and Astronomy, University of New Hampshire, USA}

\begin{abstract}
The solar wind, when measured close to 1 au, is found to flow mostly radially outward. There are, however, periods when the flow makes angles up to 15$^\circ$ away from the radial direction, both in the east-west and north-south directions. Stream interaction regions (SIRs) are a common cause of east-west flow deflections. Coronal mass ejections (CMEs) may be associated with non-radial flows in at least two different ways: 1) the deflection of the solar wind in the sheath region, especially close to the magnetic ejecta front boundary, may result in large non-radial flows, 2) the expansion of the magnetic ejecta may include a non-radial component which should be easily measured when the ejecta is crossed away from its central axis. 

In this work, we first present general statistics of non-radial solar wind flows as measured by STEREO/PLASTIC throughout the first 13 years of the mission, focusing on solar cycle variation. We then focus on the larger deflection flow angles and determine that most of these are associated with SIRs near solar minimum and with CMEs near solar maximum. However, we find no clear evidence of strongly deflected flows, as would be expected if large deflections around the magnetic ejecta or ejecta with elliptical cross-sections with large eccentricities are common. We use these results to develop a better understanding of CME expansion and the nature of magnetic ejecta, and point to shortcomings in our understanding of CMEs.

\end{abstract}



\section{Introduction}

\begin{figure}[ht]
\centering
{\includegraphics[width=13cm]{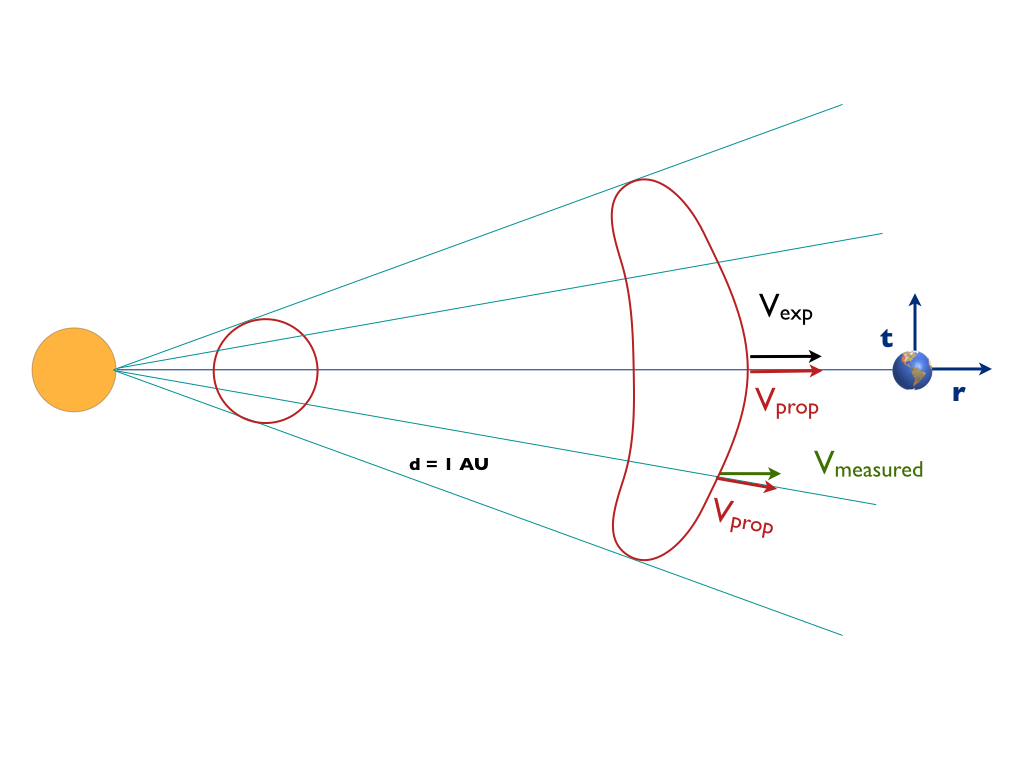}} 
\caption{Expected CME shape due to the kinematic distortion}
\label{fig:kinematic Model}
\end{figure}

Coronal Mass Ejections (CMEs) are large solar plasma eruptions that are dominated by the magnetic field. They occur frequently, especially near solar maximum conditions. They have been a subject of numerous studies since their first discovery in 1971 \citep[]{Tousey:1973}. Since their first {\it in situ} observation by five longitudinally-spaced spacecraft in a rare conjunction event in 1979 \citep[]{Burlaga:1981}, they have been considered as coherent structures. For a significant proportion of CMEs, the internal magnetic field can be adequately described as that of a twisted flux rope \citep[]{Gold:1962,Klein:1982,Burlaga:1988}. Consequently, and rather for simplicity, it was first assumed that the CME cross-section is circular. Following that, several fitting techniques to estimate the magnetic field configuration from {\it in situ} observations have been developed, the vast majority of which assume a circular cross-section \citep[]{Lepping:1990,Marubashi:1986,Farrugia:1993,Nieves:2016}. 
Elliptical cross-section has also been alternatively proposed based on the idea that, as CMEs propagate, they deform, a process commonly referred to as ``pancaking''. This was also proposed because a circular cross-section was deemed a strong simplification made for the benefit of writing down simple analytical models, to which the natural extension is an elliptical cross-section.
To further account for the dynamic changes in the CME shape during propagation, \citet{Riley:2004b} and \citet{Owens:2006b},
using a \textbf{kinematics} model, presented a scenario in which the CME starts as a flux rope with a circular cross-section that evolves radially in heliocentric coordinates. In such a scenario, the circular cross-section evolves kinetically into a convex-outwards, bean-shaped ellipse as can be seen in Figure~\ref{fig:kinematic Model}.

Such a deformation is also typically observed in numerical simulations, especially for CME models that do not have internal magnetic field, such as ENLIL \citep{Odstrcil:1999}
as well as for those initiated with flux ropes or spheromaks \citep[]{Manchester:2004a,Chane:2006,Shiota:2016,Scolini:2019}, albeit the deformation is not always as extreme. The elliptical cross-section analytical fitting technique of \citet{Hidalgo:2002} is an example of an early application of the concept of elliptical cross-sections to the fitting of {\it in situ} measurements of CMEs, and this has more recently been revisited by \citet{Nieves:2018b}. In that work, a mathematical and fitting model is developed to take into consideration the elliptical cross-section resulting from the interaction of a CME with the solar wind. Other works, such as \citet{Erdelyi:2009}, have developed models of flux ropes with different elliptical cross-sections for application to propagation of magneto-hydrodynamical waves inside flux ropes.

Statistical analyses of {\it in situ} measurements of the distribution of impact parameters (the distance between the spacecraft path and the axis of the flux rope) of more than 100 CMEs by \citet{Lepping:2010} have been used by \citet{Demoulin:2013} to deduce that the CME cross-section is elliptical with a ratio of minor (radial direction) to major (tangential or normal direction) of 1:2 or 1:3. Simultaneous measurements of magnetic ejecta (MEs) by {\it Wind} and ACE when separated longitudinally by ~0.01 au have been found to be consistent with an elliptical cross-section with a similar eccentricity \citep[]{Lugaz:2018}. \citet{Nieves:2018b} showed an example of a CME measured {\it in situ} by {\it Wind} for which the model with an elliptical cross-section fits the magnetic field measurements significantly better than a model with a circular cross-section.

The idea of elliptical cross-section has been taken further by \citet{Owens:2017}. Based on the widely accepted notion that the latitudinal extent of CMEs remain constant as they propagate \citep[e.g., see][]{Riley:2004b}, the simple kinematic model of CME propagation explained earlier \citep{Owens:2006b} results in the major axis getting significantly larger than the minor axis to the point that the latitudinal expansion speed becomes larger than the local Alfvén speed (see Figure~2 from \citet{Owens:2017}). Based on this, \citet{Owens:2017} concluded that the CME becomes an incoherent structure at a heliocentric distance of 0.2--0.5~au, depending on its (fixed) latitudinal width. 
It is worth noting that such a model is purely kinematic and does not account for the complex roles played by the various forces present in the CME (for a discussion of forces acting on CMEs as they propagate, see for example \citet{FShen:2012} and \citet{Kay:2021}). In addition, such a model also places a constraint on the evolution of CMEs, ignoring major processes that impact CME shape such as: expansion, deflection, rotation and erosion. To that effect, \citet{Suess:1988} illustrated a model of the CME evolution from a circular to an elliptical cross-section in a manner similar to that presented later by  \citet{Riley:2004b}, but highlighted how the effect of the magnetic tension must be strong enough to compensate for this deformation and to keep the cross-section nearly circular.


One would expect these studies to have investigated the presence or absence of non-radial flows as a potential signature of the evolving ejecta shape. However, very few works up until now have touched on non-radial flows associated with CMEs. For example, the model of \citet{Wang:2016} represents CMEs with a circular cross-section, and it uses the information provided by the three components of the velocity vector, along with the three components of the magnetic field vector to fit simultaneously the CME magnetic field and velocity. This includes its radial propagation, any non-radial bulk motion as well as a uniform expansion to maintain the circular cross-section. While non-radial flows are considered, the model is limited by its assumption of a circular cross-section. Another example is \citet{Owens:2004b}, who investigated the presence of large non-radial flows associated with CMEs. However, they focused on the flows associated with the CME sheath regions, as those tend to be larger than the non-radial flows inside ejecta. Quoting from their conclusion, the authors noted ``the existence of significant non-radial flows in the body of ejecta, though the magnitude of such flows are nominally less than the preceding sheath region.'' 
 Except for these works, non-radial flows inside CMEs have not been investigated or used in models to better reflect the CME elliptical cross-section in a comprehensive manner. 

 In the present work, we study non-radial plasma flows as measured near 1~au over solar cycle 24 in a comprehensive manner. In section~\ref{sec:expectations}, we lay out the different scenarios in which a CME is expected to change size and shape as it propagates and the associated flow signatures. In section~\ref{sec:statistics}, we describe the data and the methodology used in this study, the initial statistical analysis, including the average non-radial flows measured each year during solar cycle 24, and their solar cycle variations. In section~\ref{sec:causes}, we perform an analysis on the larger instances of non-radial flows. We study their causes and investigate their association with CMEs or SIRs, and also with substructures inside the CMEs (ejecta, sheath). 
 In section~\ref{sec:expandedcme}, we focus on those CMEs that have strong radial expansion flows to determine whether they are associated with non-radial flows inside the ejecta. We discuss our results and conclude in section~\ref{sec:conclusions}.

\section{Non-radial Flows: Model-Dependent Expectations}\label{sec:expectations}

While investigations of MEs have focused on the magnetic field measurements, plasma measurements provide additional information about the magnetic configuration of the ejecta, to which little, if any, attention has been devoted. In this section, we discuss why some non-radial flows are expected to be present inside those MEs which are crossed away from their center, and how their direction should give information about the ejecta shape. We present three scenarios based on past works. As CME radial expansion is well attested \citep[]{Klein:1982,Suess:1988,Farrugia:1993,Lugaz:2017}, all scenarios include the CME radial expansion. We emphasize that most of these work do not consider non-radial flows, but as illustrated below, such flows should be present unless the circumstances are close to unique, as explained in the first scenario. 

\subsection{Scenario $\#$1; Pure radial flows}\label{sec:scenario1}  

\begin{figure}[ht]
\centering
{\includegraphics[width=13cm]{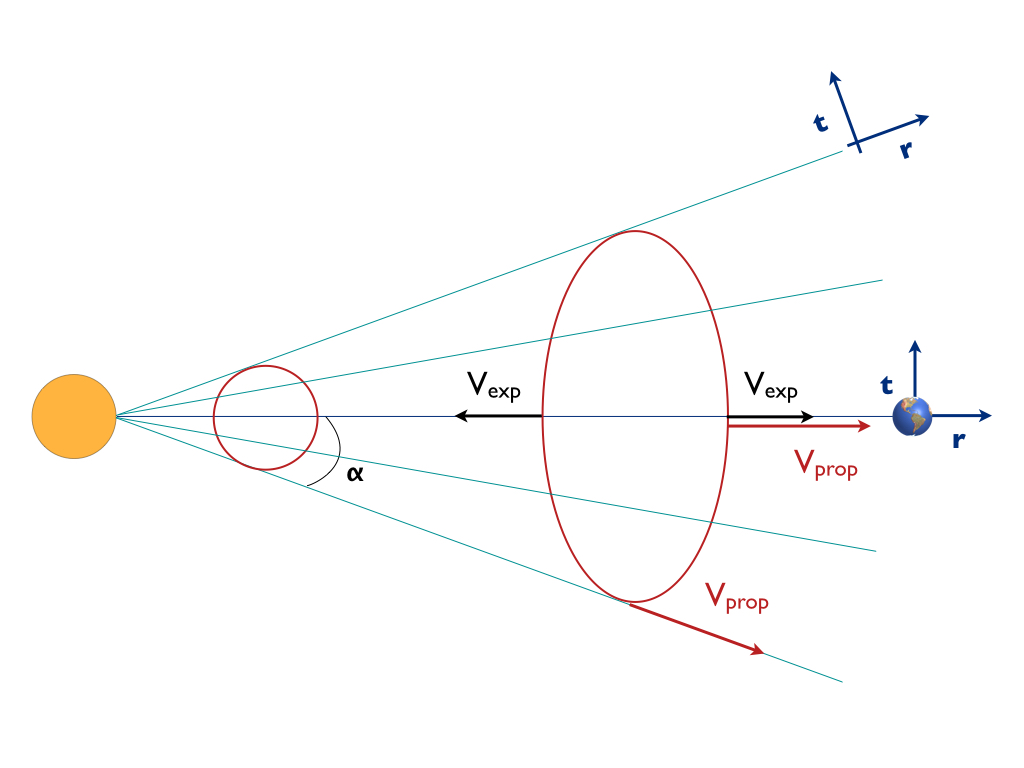}} 
\caption{Schematic to illustrate scenario $\#$1 associated with pure radial flows. The CME cross-section is shown in red, starting from being circular with a half angle of $\alpha$ and becoming elliptical further away as it maintains its angular width, 2$\alpha$. The propagation speed is $V_\mathrm{prop}$ and  $V_\mathrm{exp}$ is the expansion speed, only in the radial ($r$) direction. This schematic follows \citet{Riley:2004b} kinematics model.}
\label{fig:scenario1}
\end{figure}

This first scenario would explain the absence of non-radial flows. As the CME lateral expansion is purely kinematic, it is not associated with bulk plasma motion away from the CME axis, i.e. it is a direct consequence of the purely radial propagation of the CME and of its maintaining a constant angular width (see Figure~\ref{fig:scenario1}). However, since radial expansion and radial expansion flows are well attested, this scenario needs to explain why no expansion flows are measured in the non-radial direction. This could be because the CME magnetic tension force acts in a way that limits and hinders expansion in the non-radial direction. However, because the solar wind is radially structured, the expansion in the radial direction should be even more constrained than that in the non-radial direction. Therefore, this possibility is unlikely. Another possibility would be: in order for the CME to maintain its angular width, non-radial expansion is required. This means that the non-radial flow velocity would have to be the exact value needed to balance out the physical mechanisms that compress the CME angular width. 
This is also highly unlikely as only one value of the non-radial expansion is possible to maintain the flow exactly radial. From the examination of these two possibilities, we conclude that non-radial flows are very likely to be present inside MEs as they propagate. A similar conclusion was reached by \citet{Suess:1988}.

\subsection{Scenario$\#$2: Elliptical cross-section with large eccentricity}\label{subsec:scenario2}

First, we note that the presence of radial expansion will have the effect of making the CME cross-section more circular, unless there is non-radial expansion. This has been previously discussed by \citet{Owens:2006b} for example.
\begin{figure}[ht]
\centering
{\includegraphics[width=13cm]{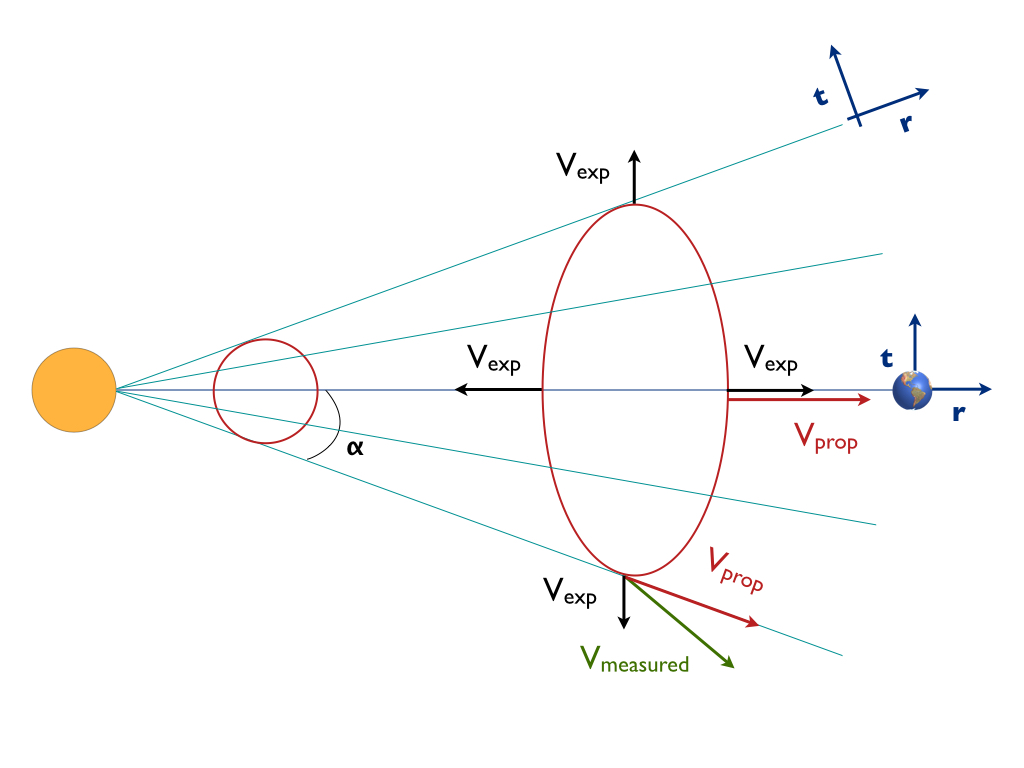}} 
\caption{Schematic to illustrate scenario $\#$2 : Adapted from \citet{Riley:2004b}'s Figure 4 (no tension) scenario.
$V_{prop}$ is assumed to be radial. $V_{exp}$ is assumed to be uniform in all directions. Actual CME shape should be similar to that discussed in \citet{Suess:1988} and \citet{Riley:2004b}. In this scenario, the tangential flow (along $t$) should be directed away from CME center/axis, resulting in non-radial flows. As such, the measured flow $V_\mathrm{measured}$ has both radial and tangential components.}
\label{fig:scenario2}
\end{figure}
This occurs because the radial expansion partially compensates the kinematically-driven lateral expansion.
However, as discussed above, radial expansion should be associated with non-radial expansion of at least similar magnitude. Because the solar wind plasma is radially stratified, the non-radial expansion might be, in fact, of larger magnitude than the radial expansion, as there is no strong pressure gradient force hindering it. Under this scenario, as illustrated in Figure~\ref{fig:scenario2}, there should be non-radial flows measured {\it in situ} when the CME is crossed away from its center. These non-radial flows are associated with the CME expansion, have a magnitude of $V_\mathrm{exp}$, and are directed {\it away} from the CME center. The net effect of the non-radial expansion combined with the kinematic expansion is to make the CME cross-section highly elliptical.

\subsection{Scenario $\#$3: Elliptical cross-section with small eccentricity or circular cross-section}\label{subsec:scenario3}

As is clear from the discussion in the previous subsections, if MEs have a circular cross-section near 1~au, there should be non-radial flows {\it towards} the ejecta center to maintain this circular cross-section. These non-radial flows are needed to compensate for the kinematic distortion of the MEs. The most likely origin would be for these flows to be associated with the magnetic tension force as discussed first by \citet{Suess:1988}. In that work, the author argues that the interaction between the CME and the solar wind results in flows away from the center in the radial direction and flows towards the center in the non-radial direction(s). These non-radial flows should only be measured when CMEs are crossed far from their center. These flows do not truly correspond to expansion but result from the difference between the expected radial propagation of all points within a CME and the measured velocity needed to keep the cross-section circular.

\begin{figure}[ht]
\centering
{\includegraphics[width=13cm]{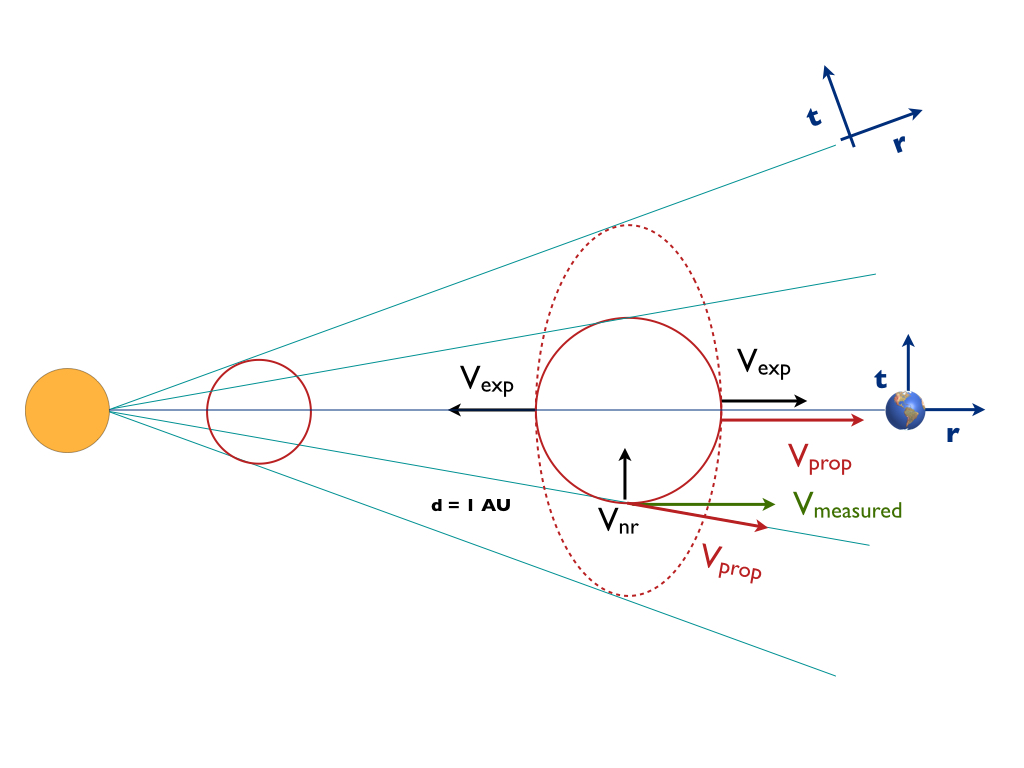}} 
\caption{Schematic to illustrate scenario $\#3$: Adapted from \citet{Suess:1988}. Magnetic ejecta are kinematically pushed to become elliptical as illustrated by the the dashed shape.
$V_{exp}$ in the radial direction compensates partially for the  kinematic effect. In  addition, tension force keeps the CME cross-section circular. The result is that points far away from the CME center are observed to propagate towards the center with a non-radial speed $V_{nr}$. Note that it would mean that the CME angular size shrinks with distance.}
\label{fig:scenario3}
\end{figure}

These non-radial flows should reach values in excess of 100--200~km\,s$^{-1}$ for the CME cross-section to remain circular (see Figure~\ref{fig:scenario3}). Here, we follow the estimate of the non-radial separation speed as computed by \citet{Owens:2017}. This is the estimate of the separation speed between two points located on opposite sides of the CME outer boundary. In that work, the authors concluded that CMEs may be incoherent structures as the non-radial separation speed becomes greater than the Alfv{\'e}n speed inside the CME.

A mixed scenario is possible for which the magnetic tension force is unable to fully counteract the kinematic distortion. However, for the CME to be less elliptical than the shape presented in scenario $\#$1, non-radial flows towards the CME center/axis are required. If such flows are not present, the CME cross-section cannot be circular and should be at least as elliptical as found by the kinematic model.

As a summary, since radial expansion is actually measured, it is nearly certain that non-radial flows should also be measured. Their direction shall inform us on a) the shape of the CME ejecta, and b) whether radial flows are associated with expansion or the restraining effect of the magnetic tension force. 
Therefore, investigating the non-radial flows as a CME is passing by a spacecraft should reveal significantly large flows in the plane perpendicular to the CME cross-section. Even if these flows do not reach values as large as those predicted by \citet{Owens:2017}, such non-radial flows should reveal a great deal of information on the cross section of the CME. In addition, absence of such flows would bring questions on the validity of the assumptions behind the above models.

\section{Study 1: Statistics}\label{sec:statistics}

\subsection{Data used and Methodology}
We use plasma data measured by the PLASTIC instrument  \citep[]{Galvin:2008} on the twin STEREO spacecraft between the years 2007 and 2019. We use STEREO-A/PLASTIC 1-minute or 10-minute data including north/south and east/west flow angles/speeds. 
The north/south flow angle is derived from deflectors on the electrostatic analyzer (ESA), which have a linear response and $\pm 20^\circ$ field of view, while the east/west flow is derived from a non-linear response position anode with a nominal $45^\circ$ field of view (which includes the velocity-dependent aberration angle). The East/West flow also has lower counting statistics. As such, the north/south flow is more accurately obtained. 

For this initial statistical study, we used plasma data at 10-minute resolution from STEREO-A only as STEREO-B was lost in 2014 and therefore the STEREO-B data do not cover a full solar cycle. Using this dataset, we have obtained the maximum and minimum flow angles and non-radial flow speeds for the whole period. In addition, the values of the following statistical quantities were obtained for each year: maximum, minimum, mean, median, standard deviation, and skewness. Data is plotted here in $RTN$ coordinates where $R$ is radially outwards from the Sun, $T$ is along the cross product of the Sun's rotation axis and the $R$ direction, and $N$ is along the $R \times T$ direction, so that the $R-N$ plane contains the Sun's rotation axis.

\subsection{Overall Statistics}

\begin{figure}[ht]
\centering
{\includegraphics[width=12cm]{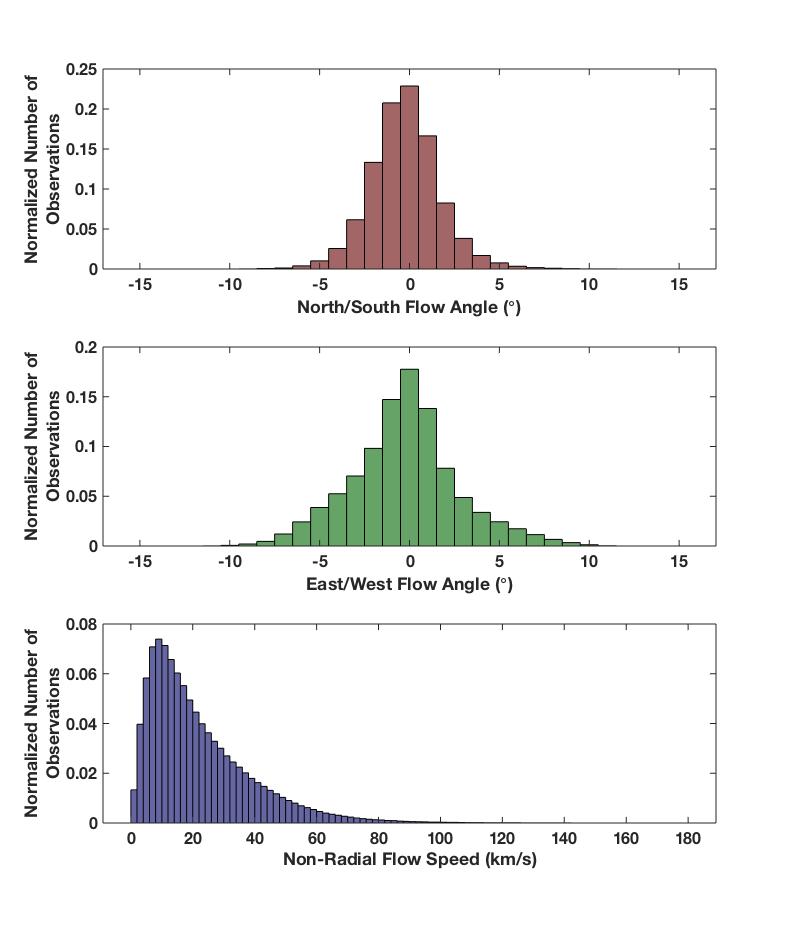}} 
\caption{Statistics of Non-Radial Flows as Measured by STEREO-A from 2007 to 2019. Top panel shows the north-south ($N$) flow angle, the middle panel the east-west ($T$) flow angle and the bottom panel the non-radial flow speed. Each panel shows the normalized occurrence numbers in bins of 1$^\circ$ for the flow angles and of 2~km\,s$^{-1}$ for the non-radial flow speed.}
\label{fig:Statistics}
\end{figure}

The statistical results are shown in Figure~\ref{fig:Statistics}. It illustrates that the large majority of measurements are made within $\pm 5^\circ$ of the radial direction and that there is no preference for a specific direction (overall symmetric profiles). This indicates that the solar wind flows primarily radially outward. However, there are some deflected flows as high as 15$^\circ$ away from radial both in the east/west and north/south directions. About $2\%$ of flows occur beyond $5^\circ$ north or south from the ecliptic and $\sim 11\%$ away from the radial direction in the east-west direction. The wider distribution of flow angles away from radial in the east-west direction rather than north-south may be related to the different ways they are measured or may be due to the fact that SIRs and CIRs are typically associated with relatively large deflections in the east-west direction (as described below). 

The typical (i.e. most common) non-radial flows have low magnitude as seen in the bottom panel of the figure, with a peak in the distribution around 10~km\,s$^{-1}$, but with a long tail going up to $\sim 180$~km\,s$^{-1}$. More than 8$\%$ of data points have non-radial flows above 50~km\,s$^{-1}$ and about 1$\%$ have flows above 80~km\,s$^{-1}$. For context, the average and median CME radial expansion in Solar Cycle 24 as measured by STEREO and reported by \citet{Jian:2018} is 62 and 57~km\,s$^{-1}$, respectively, so a non-radial flow speed of 80~km\,s$^{-1}$ is larger than the radial expansion inside most CMEs. 

Overall, this initial statistical survey indicates that relatively large non-radial flows are indeed measured. In order to determine what their origin may be, we first turn to a more detailed statistical analysis of the variation of the maximum flow with year as CMEs are more common in solar maximum and SIRs/CIRs in solar minimum. This is done to determine whether large non-radial flows occur in all phases of the solar cycle, or predominantly in solar minimum, as could be expected if large non-radial flows are associated mostly with SIRs/CIRs.

\subsection{Solar Cycle Variation}
We use 10-minute data from STEREO-A/PLASTIC and look at the yearly distribution of the non-radial flows. For each year, we determine the maximum, average and median flow angles in the east-west and north-south direction as well as the skewness of the distribution. The maximum value is noted irrespective of whether it occurs in the ``positive'' (north, east) or ``negative'' (south, west) directions. We use the flow angle rather than the flow speed as it removes any potential solar cycle variation due to the changes in the average solar wind speed. Note that in 2015, there is only limited data due to STEREO-A being turned off during the superior solar conjunction. 

\begin{figure}[ht]
\centering
{\includegraphics[width=8.6cm]{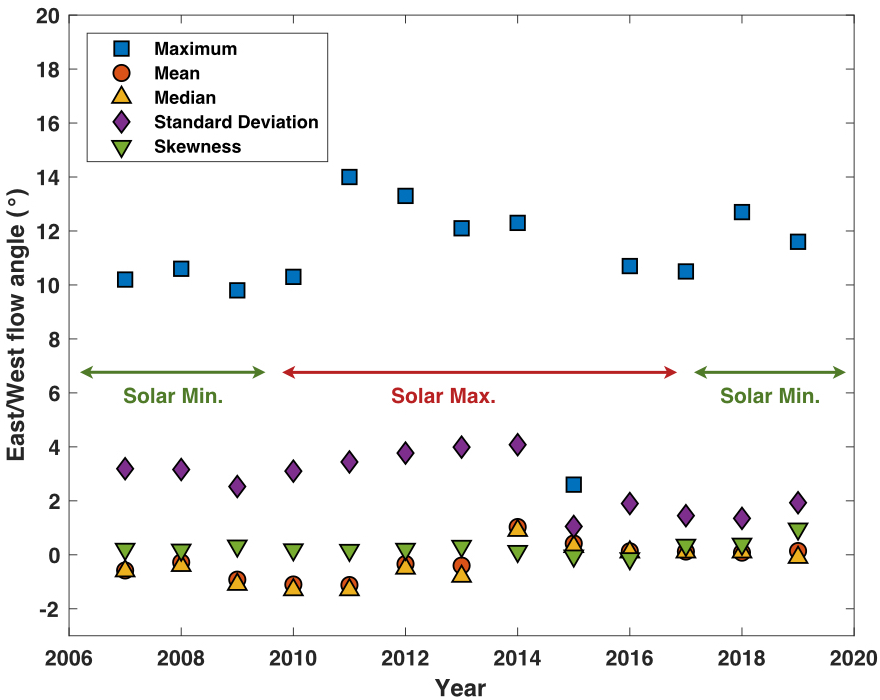}} 
{\includegraphics[width=8.6cm]{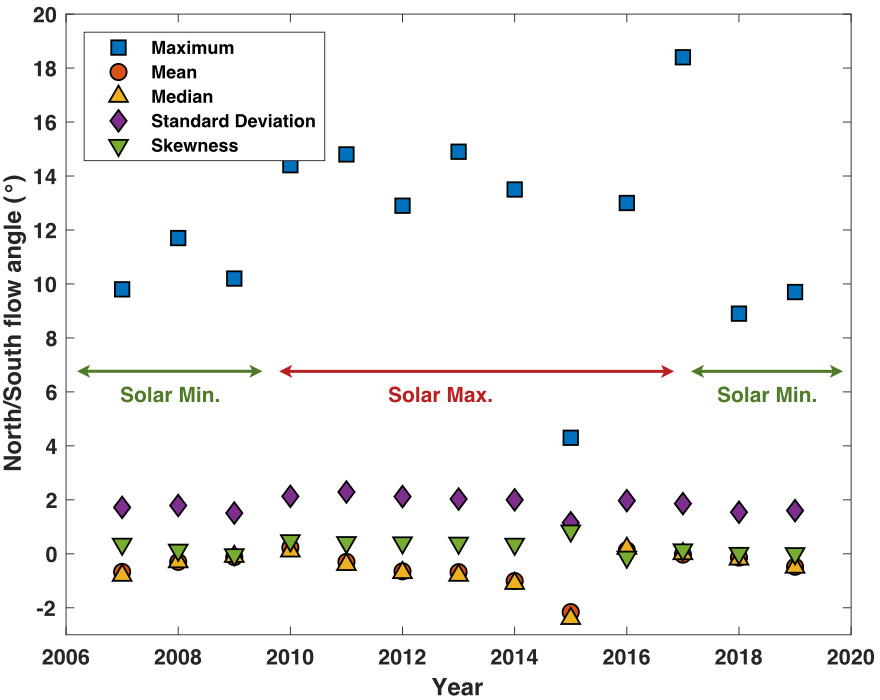}} 
\caption{This figure shows the deflection flow angles in the East/West and North/South for every year (using 1-min data). The various symbols are explained in the labels at the top of each panel. The figure shows a clear solar-cycle dependence of the maximum deflection angle. 
This indicates that CMEs which are responsible for the peak in solar maximum can be associated with the largest deflection flows. Here, solar maximum refers to the years from 2010 to 2016, including the rising and declining phases (year 2015 was a year with little data due to superior solar conjunction).
}
\label{fig:SolarCycle}
\end{figure}

Figure~\ref{fig:SolarCycle} shows the results. There is a clear solar cycle dependency in the maximum deflection flow angle measured every year. The maximum deflection flow angles are larger close to solar maximum (2011-2015) than in solar minimum and the rising/declining phase of the cycle (2007-2009, 2018-2019). 2010 and 2017 represent the two transition years between large sunspot number (and CMEs) in solar maximum and low numbers (average yearly sunspot number less than 10) in solar minimum. While not conclusive, this solar cycle variation gives us a hint that CMEs, which are much more common in solar maximum, may be associated with the largest deflection flows measured in the solar wind. In addition, CMEs were almost absent during the deep solar minimum of 2007-2009, whereas there was a significant number of SIRs. The maximum deflection flows in 2007-2009 of 10--12$^\circ$ may therefore represent the maximum deflection made possible by a SIR/CIR. In comparison, maximum deflection flows reached 13--18$^\circ$ during the maximum phase of the solar cycle, and those flows may be more likely to be associated with CMEs. We look into large flows in more details in the following section.

\section{Study 2: Causes of Large Deflection Flows} \label{sec:causes}

\begin{figure}[t]
\centering
{\includegraphics[width=10cm]{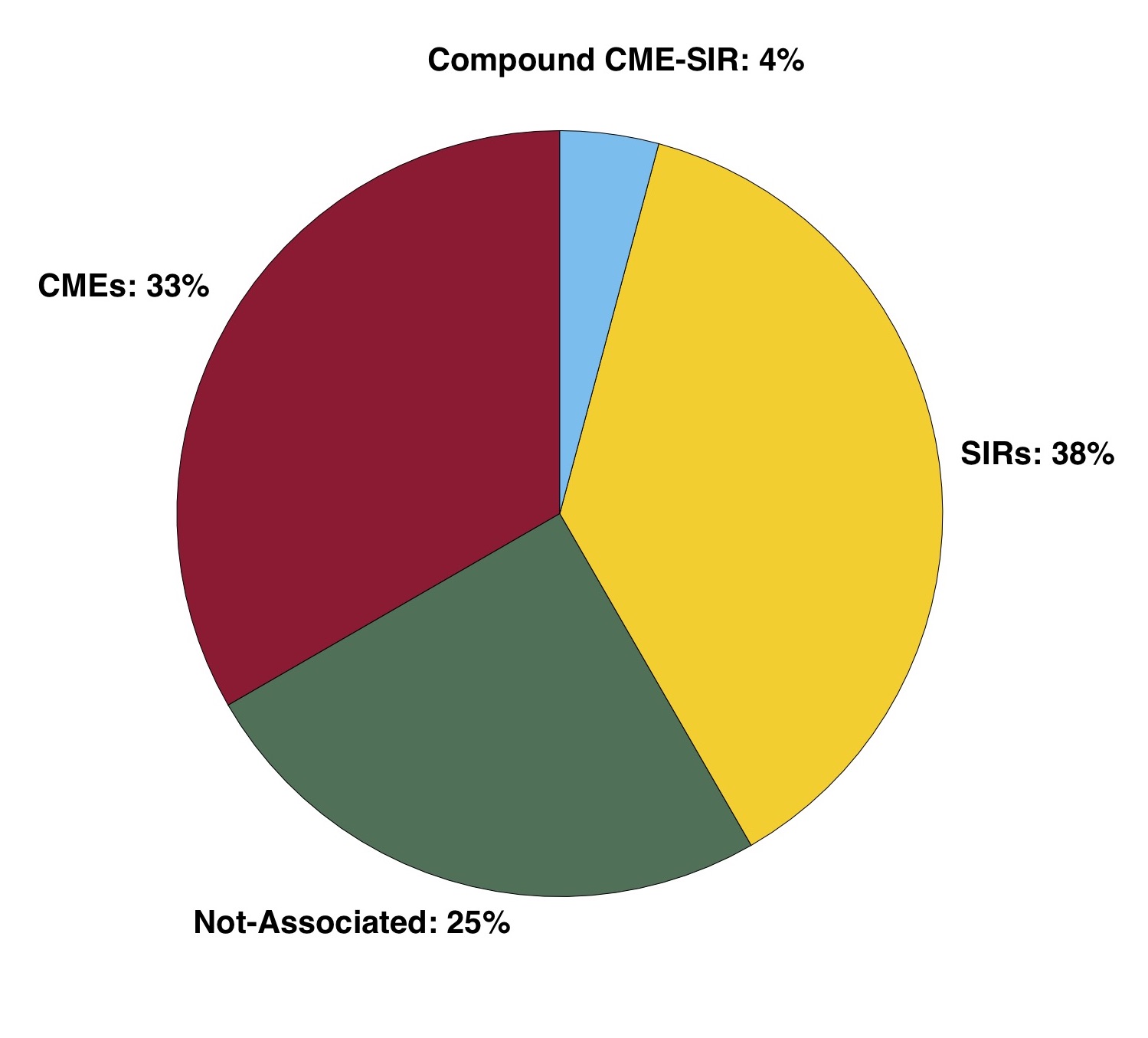}} 
\vspace{-0.2cm}
\caption{Drivers of the largest deflection flow angles (flow angle away from radial $>$ 12$^\circ$) as measured by STEREO-A from 2007 to 2019. See text for details.}
\label{fig:piechart1}
\end{figure}

Next, we identify the origin of all large non-radial flows (as characterized by Study 1) in terms of associated structures (SIRs, CMEs or others). We focus on the 24 events for which the 10-minute flow angle is greater than 12$^\circ$ either in the East/West or North/South direction for STEREO-A. 12$^\circ$ has been chosen as a threshold for large flows based on the discussion in the previous section. Since there was no deflection angle of this magnitude in 2007-2009, non-radial flows of this magnitude are likely to be associated with CMEs. We use the CMEs and SIRs database of \citet{Jian:2018} and \citet{Jian:2019} respectively, to determine the association of these flows with specific structures. The proportion of the different categories are shown in Figure~\ref{fig:piechart1} and are as follows: nine events correspond to CMEs, one of which is a compound events with an SIR, ten events correspond to SIRs, one of which is the same compound event as mentioned before, six events are not associated with CMEs nor SIRs (sometimes because of data gap). The compound event corresponds to an event that has characteristics of a CME embedded within a SIR, an occurrence which is relatively frequent. Out of the six events not associated with CMEs nor SIRs, two of them occur in the trailing edge of a CME (within 4 hours of the CME end time). Less than half of the largest flows are in fact associated with CMEs. We then turn our attention to these cases and determine the circumstances.

\subsection{Which CME sub-structure causes large non-radial flows?}

\begin{figure}[ht]
\centering
{\includegraphics[width=11cm]{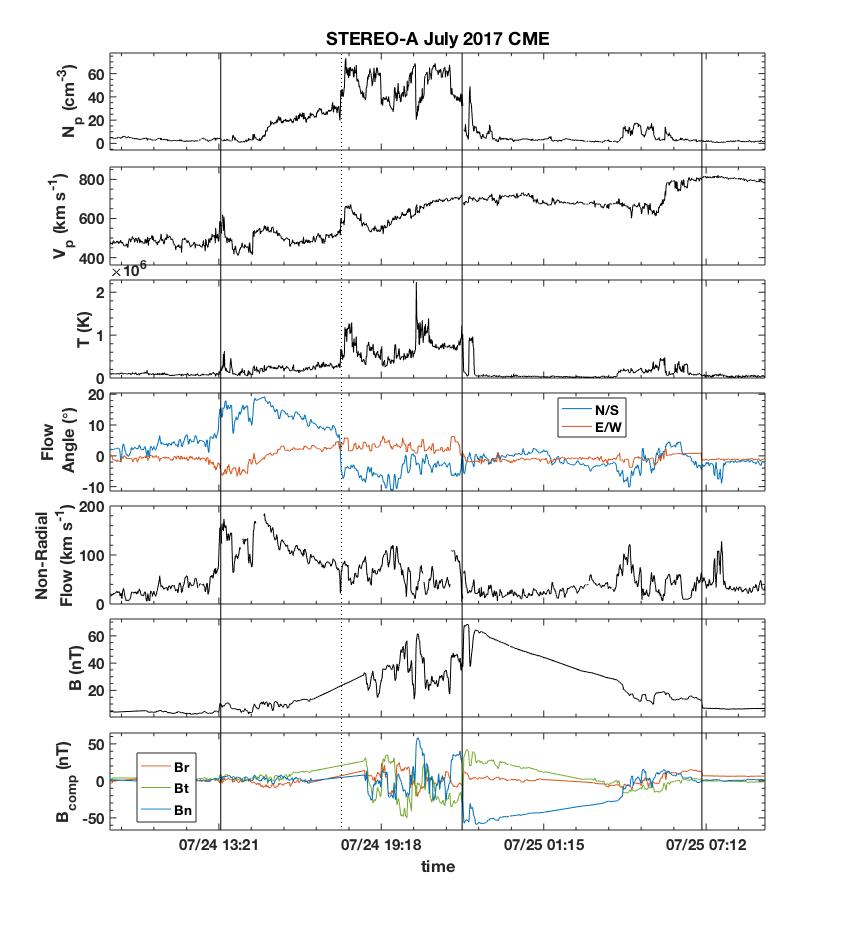}} 
\caption{July 2017 CME event measured by STEREO-A and associated with the largest non-radial flow speed angle. The panels show the proton density, velocity, temperature, the north-south (blue) and east-west (red) flow angles, the non-radial flow speed, the magnetic field magnitude and the three components of the magnetic field vector (red: radial, green: tangential, blue: normal), from top to bottom. The solid vertical lines indicate the start of the CME sheath, the start and end of the ejecta, respectively. The region of large deflection flows occur in the sheath ahead of the magnetic ejecta with a reversal of the flow direction at the dotted line from northward to southward.}
\label{fig:July17}
\end{figure}

We first turn our attention to the maximum non-radial flow angle related to a CME, that occurred between 2007 and 2019. This corresponds to the 2017 July 24 event with a flow with an angle away from radial of 18.4$^\circ$. Measurements by STEREO-A are shown in Figure~\ref{fig:July17}. This is a CME with a strong magnetic field reaching above 60 nT at the back of the sheath and the beginning of the ME. There is a forward fast magnetosonic shock at 14:36 UT on July 24, and the ejecta starts around 22:40 on July 24 and the strong magnetic field region lasts until ~7:00 UT on July 25 although the CME end time is given as later in the database of \citet{Jian:2018}. The CME propagates through moderately fast solar wind and the sheath region is relatively complicated. The large non-radial flows (blue line in the fourth panel) occur in the sheath region, with first strong northward flows until about 17:00 UT and then large southward flows (transition indicated by the dotted vertical line). These are consistent with draping (i.e. the wrapping of the magnetic field lines around the ejecta, deflecting the plasma to flow tangential to the ejecta) as discussed by \citet{McComas:1989}, for example. The two distinct deflection flow patterns are separated by a discontinuity where there are also temperature and density increases. There is no large deflection in the ME.

We next quickly consider the largest non-radial flow speed that occurs inside a CME (non plotted here). These are non-radial flow speed values of 234 km\,s$^{-1}$ associated with the 6 June 2011 CME measured at STEREO-A. This CME had a reported expansion speed of $\sim$ 100 km\,s$^{-1}$. We identify the part of the CME where these large non-radial flows occurred. They correspond to deflection in the CME sheath associated with a complex event (potentially two separate flux ropes). There was a very long sheath region, where deflection was large. Overall, focusing on these two most extreme events, we confirm that, indeed CME sheaths are associated with the largest deflection flows, confirming past studies such as that of \citet{Owens:2004b}. The question remains whether or not large deflection flows can be observed within the ME.

\section{Study 3: CMEs with Large Radial Expansion} \label{sec:expandedcme}
\subsection{Data Selection and Methodology}
We now turn our attention to CMEs with relatively large radial expansion speed to determine whether or not any of them are associated with non-radial expansion flows.
To perform this study, we look at all CMEs measured {\it in situ} by STEREO from 2007 to 2019 with radial expansion $>$ 100~km\,s$^{-1}$. We use 1-min PLASTIC data and the database of \citet{Jian:2018}. For each event with radial expansion speed $>$ 100~km\,s$^{-1}$, we calculate the average non-radial flow speed inside the ME (excluding the sheath) as well as the east-west and north-south flows. To do so, we average the non-radial flow speed over the duration of the ME as provided from the database of \citet{Jian:2018} (where the ME is referred to as ``magnetic obstacle''). We find 28 CME events for which the expansion speed is greater than 100~km\,s$^{-1}$ (excluding the 2012 July 23 event at STEREO-A due to the extreme speed and significant data gaps). We use this criteria of radial expansion speed $>$100 ~km\,s$^{-1}$ for the following reasons: If non-radial flows are associated with CME expansion, then they should be easier to measure when the CME has large radial expansion. Based on the study by \citet{Demoulin:2013}, the median impact parameter inside magnetic clouds is $\sim 0.3$. If such clouds are crossed at an impact parameter of 0.3 or above, while expanding uniformly at speeds greater than 100~km\,s$^{-1}$, then the associated non-radial flows should be measurable. As we find 28 events satisfying this criterion, we can expect some CMEs to be crossed farther away from their center than 0.3 (the average), which is a requirement to measure non-radial flows under scenarios 2 and 3 as detailed in section~\ref{sec:expectations}. 

\subsection{Statistics/Results}
The average magnitude of the non-radial flow speeds over these 28 events is 40~km\,s$^{-1}$ (median of 36~km\,s$^{-1}$). This can be compared to the average radial expansion of 122~km\,s$^{-1}$ (median of 108~km\,s$^{-1}$), since both of these flows can be thought as occurring in the frame of the radially propagating ME. The ratio of non-radial to radial flows in the ME frame is about one third. Non-radial flows occur slightly more in the tangential direction (average of 21~km\,s$^{-1}$, median of 13~km\,s$^{-1}$) than normal direction (average of 9~km\,s$^{-1}$, median of 7~km\,s$^{-1}$). There is no preferred north-south direction, but a slight preference for eastward flow (average of $-8$~km\,s$^{-1}$).

The largest non-radial flows reach $\sim 70$~km\,s$^{-1}$ or about 60$\%$ of the average radial expansion flows measured inside CMEs. As such, even though the magnitude of non-radial flows is less than that of radial flows associated with expansion in the ME frame, they are non-negligible. However, to determine whether or not these non-radial flows play a role in CME expansion and shape, it is necessary to investigate their direction (towards or away from the CME axis/center) in the frame of the ME, as explained through the various scenarios in section 2.

\subsection{Case Studies}
We focus on one case with relatively average non-radial flows as well as one additional case with large non-radial flows.

\subsubsection{2010 September 11 CME}

\begin{figure}[ht]
\centering
{\includegraphics[width=11cm]{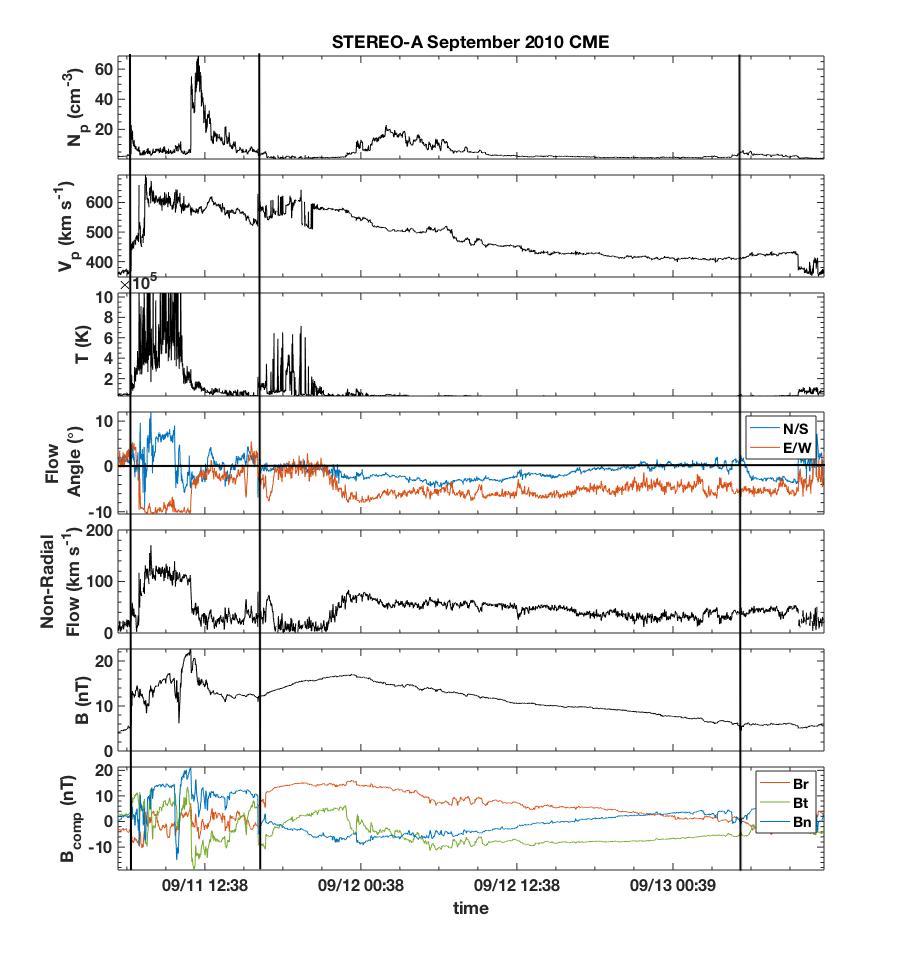}}
{\includegraphics[width=6.5cm]{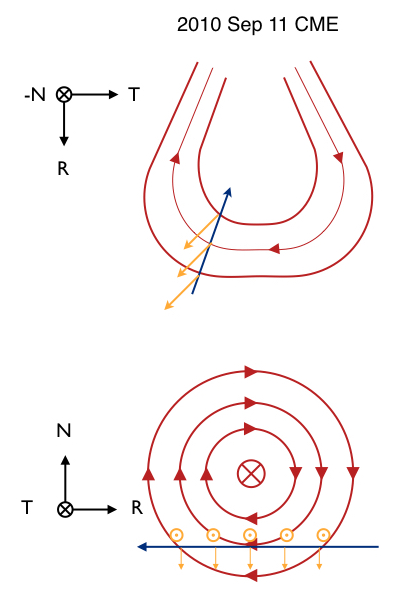}} 
\caption{STEREO-A measurements of the 2010 September 11 CME. The panels show from top to bottom, the proton density, velocity, temperature, the non-radial flow angles and the non-radial flow speed, the magnitude and components of the magnetic field.
Right: Scenario reflecting the measurements. The magnetic field is in blue, the plasma flows in orange and the green arrow indicates the proposed crossing direction of the spacecraft.}
\label{fig:Sep2010}
\end{figure}

Figure~\ref{fig:Sep2010} shows STEREO-A measurements of the 2010 September 11-13 CME. This is a relatively typical CME preceded by a dense sheath region and fast forward shock at 6:59 UT on September 11. The ME lasted over 36~hours and was crossed at a high impact parameter (notice the large $B_R$ values in last panel). The radial expansion is $\sim$ 110 km\,s$^{-1}$ as calculated from the decreasing speed profile in the second panel. The average of the non-radial flows inside the ejecta are $\sim$ 40~km\,s$^{-1}$, primarily in the east direction ($-T$) but with a component in the south direction ($-N$) in the center of the ejecta (peak of $\sim-30$~km\,s$^{-1}$ around 07:00 UT on September 12). These non-radial flows reach $\sim 75$~km\,s$^{-1}$ in the front half of the ejecta (note that the very front part of the ejecta has almost zero non-radial flows). While there are even larger non-radial flows inside the front part of the sheath, the non-radial flows inside the ejecta are noticeable in terms of magnitude but also consistent direction and decreasing magnitude as the CME passes over the spacecraft.

Next, we turn our attention to the orientation of the ME and the direction of these non-radial flows with respect to ejecta to shed light on their relation to the three scenarios discussed early on.

A fit with the circular cross-section force-free model of \citet{Lepping:1990} is performed for the magnetic field measurements from September 11 16:45 UT to September 13 05:55 UT. This model was chosen because the requirements to use the static force-free model appear to be satisfied. While more complex models exist, we intend to confirm the overall orientation (low-lying or highly inclined) and whether the crossing occurs at small or large impact parameter. For this purpose, the well validated model of \citet{Lepping:1990} is appropriate. 
The orientation is found to be low-inclined with ($\theta$, $\phi$) = ($-23^\circ$, $254^\circ$), the latitude and longitude angles of the flux rope axis with a negative chirality. As such, this is a south-west-north (SWN) cloud ($B_N$ varying from negative to positive, while $B_T$ remains negative). The impact parameter is confirmed to be high at 0.75 with $B_R$ positive especially in the front part of the ejecta. For such a low-inclined ejecta crossed at high impact parameter, the non-radial flows associated with the ejecta cross-section are expected to occur in the north-south direction (see the right panel of Figure~\ref{fig:Sep2010}). The south-directed non-radial flows are consistent with these and following scenario~2, but their magnitude are only up to 25$\%$ of the radial expansion flows even though the impact parameter is large. The dominant non-radial flows in the $-T$ direction could be associated with a global expansion of the CME axis (not its cross-section).

\subsubsection{2012 July 11 CME}
\begin{figure}[ht]
\centering
{\includegraphics[width=10cm]{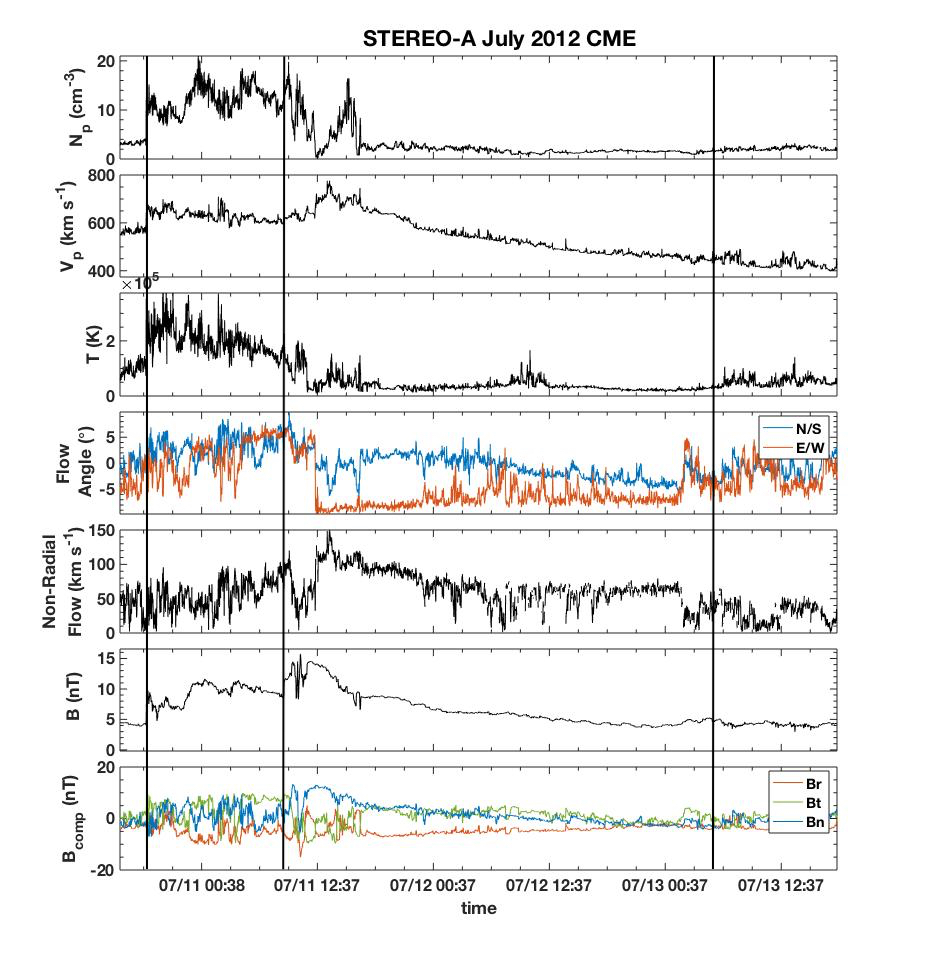}}
{\includegraphics[width=7cm]{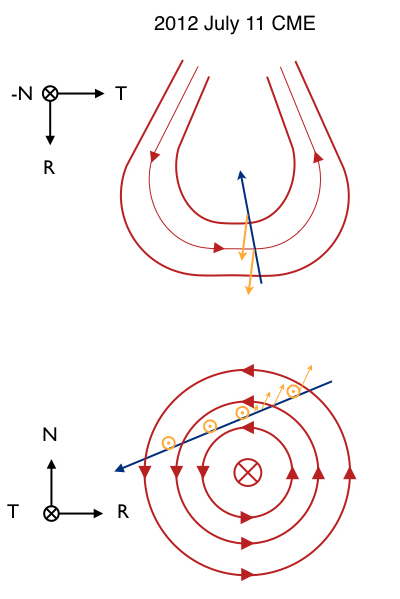}} 
 
\caption{Left: STEREO-A measurements of the 2012 July 11 coronal mass ejection. Same format as Figure~9. Right: Scenario reflecting the measurements. The magnetic field is in blue, the plasma flows in orange and the green arrow indicates the proposed crossing direction of the spacecraft. The large non-radial flows in the $-T$ direction cannot be simply explained.}
\label{fig:Jul2012}
\end{figure}

Figure~\ref{fig:Jul2012} shows STEREO-A measurements of the 2012 July 11 CME. The flow makes an angle of $\sim 8-10^\circ$ in the east direction throughout the ME with an additional deflection by $\sim 3^\circ$ towards the north in the early part of the ejecta. Taken together, this corresponds to non-radial speeds reaching above 120~km\,s$^{-1}$, which are comparable in magnitude to the radial expansion of the ejecta. As for the previous event, the radial component of the magnetic field is relatively elevated throughout the ME possibly indicating a crossing away from the ME center, with $B_R$ being especially large early on in the ejecta and decreasing in magnitude.

The magnetic field within the ME is not well-organized, but a fitting is possible from 17:00 UT on July 11 to 06:00 UT on July 13 with the force-free circular cross-section model of \citet{Lepping:1990}. The orientation is found to be low-inclined with ($\theta$, $\phi$) = ($-20^\circ$, $94^\circ$), the latitude and longitude angles of the flux rope axis with a negative chirality. As such, this is a north-east-south (NES) cloud ($B_N$ varying from positive to negative, while $B_T$ remains positive). The impact parameter is confirmed to be high at 0.79. For such a low-inclined ejecta crossed at high impact parameter, the non-radial flows are expected to occur in the north-south direction (see the bottom right panel of Figure~\ref{fig:Jul2012}). Since $B_R$ is large at first, the initial flows in the northern direction indicate flows away from the center, consistent with scenario 2. However, even with a large radial expansion and a large impact parameter, the north-south flows are only a small fraction ($\sim$ 20\%) of the radial expansion. The non-radial flows in the $-T$ direction cannot easily be understood except as a general deflection of the ejecta, or as a sign that the CME magnetic morphology is more complex than one with invariance along a central axis.

\section{Discussion and conclusions}\label{sec:conclusions}

Our expectations are that there should be significant ($> 5^\circ$) non-radial flows measured in association with the crossing of CME ejecta away from their center (high impact parameters). These flows should occur in all the following cases: (a) if the CME expansion, which is measured in the radial direction, is isotropic, (b) if the CME expansion occurs faster in the non-radial direction, (c) if the magnetic field tension force keeps the cross-section circular. 

In the current study, we first looked at all statistics of non-radial flows as measured by STEREO/PLASTIC over the first 13 years of its mission. We determined that non-radial flows that are comparable in magnitude to large radial expansion flows do occur. We then turned our attention to the solar cycle variation, finding clear signs that maximum values of non-radial flow angles occur close to solar maximum, whereas solar minimum values do not go beyond 10-12$^\circ$ away from the radial direction. This indicates that CMEs, that are more common during solar maximum, may be associated with the largest flows.

We then turned our attention to the instances where non-radial flow angles reached values of 12$^\circ$ or more and found that $\sim 40\%$ were indeed associated with CMEs, about the same number with SIRs, and the rest not clearly associated with any transients. We then confirmed that the largest non-radial flows inside CMEs are associated with deflections in the sheaths.

In the last part of this study, we looked at the magnitude of non-radial flows for CMEs with large radial expansion speeds measured by STEREO. We determined that the average non-radial flow speeds inside 28 such events was 40~km\,s$^{-1}$ or about one third of their radial expansion speed. We looked at two events to determine in which directions these flows are with respect to the ejecta orientation. In both cases, the flows occur primarily along the flux rope axis and not in the direction of the CME cross-section. In both cases, there are small flows consistent with small non-radial expansion, which would imply that the CME cross-section is somewhat more elliptical than what is implied from kinematic models such as that of \citet{Riley:2004b}. The presence of large non-radial flows in the direction of the CME axis cannot be easily explained. While in one of our case studies it could be understood as an expansion of the CME axis, in the other case, it is in the opposite direction. It might be associated with an overall deflection of the CME, although the absence of such flows in part of the sheath for the 2010 September 11 CME would mean that the sheath and the ME are not deflected in the same way. Instrument effects may be at play, especially since flows along the normal direction (north-south) are measured through deflectors whereas flows along the tangential direction (east-west) are derived from impacts on the microchannel plates (MCPs). 

Overall, a surprising result of this investigation is that there are no steady large non-radial flows inside CME ejecta, as would be expected from uniform expansion or the effect of the magnetic tension force. As discussed in section~\ref{sec:expectations}, this can only be understood if the radial expansion is not associated with an expansion in the non-radial direction and if the CME cross-section becomes highly elliptical due to kinematic effects. Alternatively, the entire paradigm of flux rope with an axial invariance might be too limited (see past discussion in \citet{AlHaddad:2011}). Multi-spacecraft measurements that include plasma measurements may provide an opportunity to further investigate this problem. They will be more likely in the near future as STEREO returns to the vicinity of the Sun-Earth line while solar activity picks up. However, such multi-spacecraft measurements may not always be able to help distinguish between different CME structures \citep[]{AlHaddad:2019}.

\begin{acknowledgments}
This research was supported by NASA grants 80NSSC21K0463, 80NSSC20K0431, 80NSSC17K0556 and NSF grants AGS1954983.
\end{acknowledgments}

\bibliographystyle{aasjournal}

\begin{table}[]
\centering
\begin{tabular}{|l|l|l|l|l|l|l|l|l|l|l|l|l|l|}
\hline
\multicolumn{14}{|c|}{East\_West flows}                                                                                    \\ \hline
Year               & 2007  & 2008  & 2009  & 2010  & 2011  & 2012  & 2013  & 2014  & 2015  & 2016  & 2017  & 2018  & 2019  \\ \hline
Maximum            & 10.2  & 10.2  & 9.4   & 9.9   & 14    & 13.3  & 12.1  & 12.3  & 2.6   & 9.7   & 10.2  & 10.8  & 11.6  \\ \hline
Minimum            & -9.9  & -10.6 & -9.8  & -10.3 & -10.9 & -11.1 & -10.4 & -10.7 & -2.3  & -10.7 & -10.5 & -12.7 & -9.2  \\ \hline
Mean               & -0.58 & -0.28 & -0.92 & -1.1  & -1.12 & -0.34 & -0.4  & 1.03  & 0.42  & 0.13  & 0.12  & 0.08  & 0.14  \\ \hline
Median             & -0.6  & -0.4  & -1.1  & -1.3  & -1.3  & -0.5  & -0.8  & 0.9   & 0.3   & 0.1   & 0.1   & 0.1   & -0.1  \\ \hline
Standard deviation & 3.19  & 3.16  & 2.53  & 3.1   & 3.44  & 3.77  & 3.99  & 4.08  & 1.05  & 1.9   & 1.45  & 1.35  & 1.93  \\ \hline
Skewness           & 0.21  & 0.18  & 0.33  & 0.19  & 0.17  & 0.21  & 0.32  & 0.13  & -0.04 & -0.14 & 0.36  & 0.4   & 0.96  \\ \hline
\multicolumn{14}{|c|}{North-South   flows}                                                                                 \\ \hline
Maximum            & 9.8   & 11.7  & 10.2  & 14.4  & 14.8  & 10.9  & 14.9  & 13.5  & 2.5   & 13    & 18.4  & 8.4   & 7.8   \\ \hline
Minimum            & -9.1  & -10.7 & -9.3  & -11.3 & -13.4 & -12.9 & -10.5 & -10.2 & -4.3  & -10.7 & -13.8 & -8.9  & -9.7  \\ \hline
Mean               & -0.67 & -0.3  & -0.12 & 0.23  & -0.3  & -0.65 & -0.68 & -1.01 & -2.16 & 0.13  & -0.03 & -0.14 & -0.48 \\ \hline
Median             & -0.8  & -0.3  & -0.1  & 0.1   & -0.4  & -0.7  & -0.8  & -1.1  & -2.4  & 0.2   & 0     & -0.2  & -0.5  \\ \hline
Standard deviation & 1.72  & 1.79  & 1.51  & 2.13  & 2.29  & 2.12  & 2.03  & 2     & 1.15  & 1.97  & 1.86  & 1.54  & 1.6   \\ \hline
Skewness           & 0.36  & 0.14  & -0.03 & 0.49  & 0.42  & 0.41  & 0.4   & 0.36  & 0.85  & -0.12 & 0.15  & 0.02  & 0.01  \\ \hline
\end{tabular}
\caption{}
\label{tab:my-table}
\end{table}



\begin{table}[]
\centering
\begin{tabular}{|l|l|l|l|}
\hline
\textbf{Year} &\textbf{value of flow} & \textbf{Event Type} & \textbf{Date of Flow} \\ \hline
\textbf{2017} & -13.8  & CME       & 22-Oct-2017 7:40:39  \\ \hline
\textbf{2011} & -13.4  & CME       & 22-Mar-2011 19:40:23 \\ \hline
\textbf{2012} & -12.9  & None      & 29-May-2012 6:20:01  \\ \hline
\textbf{2011} & -12.80 & CME       & 10-Sep-2011 9:00:10  \\ \hline
\textbf{2018} & -12.7  & None      & 2-Nov-2018 23:20:08  \\ \hline
\textbf{2011} & -12.40 & CME       & 6-Jun-2011 3:40:16   \\ \hline
\textbf{2013} & 12.1   & CME       & 3-Mar-2013 14:50:37  \\ \hline
\textbf{2014} & 12.20  & None      & 27-May-2014 12:40:44 \\ \hline
\textbf{2014} & 12.3   & None      & 27-Jan-2014 15:40:01 \\ \hline
\textbf{2011} & 12.30  & SIR       & 21-May-2011 11:20:17 \\ \hline
\textbf{2011} & 12.50  & SIR       & 14-Mar-2011 22:30:23 \\ \hline
\textbf{2012} & 12.70  & None      & 18-Jun-2012 19:50:00 \\ \hline
\textbf{2011} & 12.80  & None      & 12-Apr-2011 14:20:20 \\ \hline
\textbf{2016} & 13     & SIR       & 13-Nov-2016 10:30:58 \\ \hline
\textbf{2012} & 13.3   & CME       & 16-Jul-2012 16:30:58 \\ \hline
\textbf{2014} & 13.5   & CME + SIR & 16-Feb-2014 8:30:57  \\ \hline
\textbf{2011} & 13.50  & SIR       & 9-Jul-2011 19:30:15  \\ \hline
\textbf{2011} & 13.90  & SIR       & 4-Jun-2011 15:20:16  \\ \hline
\textbf{2011} & 14     & SIR       & 26-Jul-2011 8:40:14  \\ \hline
\textbf{2013} & 14.20  & SIR       & 3-Jul-2013 20:20:27  \\ \hline
\textbf{2010} & 14.4   & SIR       & 8-May-2010 4:20:55   \\ \hline
\textbf{2011} & 14.8   & SIR       & 16-Jun-2011 22:50:15 \\ \hline
\textbf{2013} & 14.9   & CME       & 2-Mar-2013 5:00:37   \\ \hline
\textbf{2017} & 18.4   & CME       & 24-Jul-2017 14:40:42 \\ \hline
\end{tabular}
\caption{}
\label{tab:my-table}
\end{table}

\end{document}